\documentclass[sigconf]{nimeart}

\usepackage{url}
\usepackage{dblfloatfix}
\usepackage{CJKutf8}
\usepackage{float}
\usepackage{hyperref}

\AtBeginDocument{%
  }

\setcopyright{cc}
\copyrightyear{2026}
\acmYear{2026}
\acmDOI{}

\acmConference[NIME '26]{International Conference on New Interfaces for Musical Expression}{June 23--26,
  2026}{London, UK}
\acmISBN{}

\begin{document}

\title{Calliphony: A Calligraphy-Driven Interface for Real-Time Generative Music Performance}

\author{Tristan WU}
\authornote{Both authors contributed equally to this research.}
\email{wwu252@connect.hkustgz.edu.cn}
\affiliation{%
  \institution{The Hong Kong University of Science and Technology (Guangzhou)}
  \city{Guangzhou}
  \country{China}
}

\author{Ruiji YU}
\authornotemark[1]
\email{ruiji.yu@mbzuai.ac.ae}
\affiliation{%
  \institution{Mohamed Bin Zayed University of Artificial Intelligence}
  \city{Abu Dhabi}
  \country{United Arab Emirates}
}

\author{Gus XIA}
\email{gus.xia@mbzuai.ac.ae}
\affiliation{%
  \institution{Mohamed Bin Zayed University of Artificial Intelligence}
  \city{Abu Dhabi}
  \country{United Arab Emirates}
}

\renewcommand{\shortauthors}{Wu et al.}

\begin{abstract}
While music generative models have recently gained significant attention, how they can be effectively integrated into live music performances still requires further exploration.
This paper presents \textbf{Calliphony}, a calligraphy-driven interface for real-time generative music performance. 
Specifically, we build a low-latency pipeline that captures brush motion with an attachable sensor 
and maps it to control signals for real-time symbolic music generation. 
Using a generative model, the system produces multi-track MIDI in performance settings, while brush-derived control signals constrain event timing and activate additional musical layers.
The generated melody is then extended with real-time harmony and additional voices, and finally rendered through a DAW for live staging.

Calliphony contributes: (1) a performance-oriented prototype that uses calligraphic motion as an external control layer for a real-time symbolic music generation model, controlling note density, pitch constraints, and accompaniment-layer activation; and (2) a cross-modal performance scenario that extends calligraphy beyond a primarily visual practice into an audiovisual, AI-assisted setting.

\end{abstract}


\maketitle

\section{Introduction}

Calligraphy is not merely a method of writing, but a long-standing visual and performative art practice~\cite{unesco2009_calligraphy,delbanco2008_chinese_calligraphy}. Many calligraphy artists think that different elements, such as paper, ink, and brushes, form an embodied system of expression: subtle dynamics of pressure, speed, pauses, and turns are inseparable from the final static written form "glyph". This makes calligraphy a compelling entry point for exploring contemporary multimedia performance and interface art~\cite{smithsonian2026_four_treasures,hotson2019_four_treasures,hung2021_calligraphy_body}. We argue that, the calligraphy writing shares a very similar "inner connection" with musical performance, where expressive meaning emerges through time via rhythm, phrasing, cadence, and silence. So, the "co-performance" for music and calligraphy emerges intuitively.   

In NIME and the related communities, researchers have combined calligraphy with music in many practices. Early work mapped brushstrokes and writing parameters to musical control signals, framing calligraphy as a gestural input modality for "writing-to-sounding" interactive systems~\cite{calligraphy_interface,calligraphy_brush}. More recent projects and discussions have shifted toward treating calligraphy as a cross-media practice, focusing on the embodied act of writing and its role in live artistic performance~\cite{phantom_utopia,calligraphy_collaboration}. Notably, a growing trend has been to incorporate AI techniques into the interaction design. However, relatively less work has explored calligraphic motion as a culturally situated, non-musical gestural interface for controlling real-time symbolic music generation in live performance.



Meanwhile, the ability of generative models to represent and generate musical structure has evolved rapidly with recent model architectures~\cite{roberts2018musicvae,yang2019deep,mittal2021symbolicdiffusion,min2023polyffusion,wang2024whole,huang2018musictransformer,dhariwal2020jukebox,agostinelli2023musiclm,copet2023musicgen}. Many existing control paradigms rely on text prompts, audio prompts, or audio-based conditioning~\cite{agostinelli2023musiclm,copet2023musicgen,dhariwal2020jukebox}, which are typically better suited to offline generation than to continuous, low-latency control in live performance. While some systems have moved toward real-time interaction—such as PerformanceRNN~\cite{Performance} and Notochord~\cite{notochord}—there is still substantial space to explore how such models can be embedded into performance settings with control schemes that are both effective and artistically novel.



Our work, \textbf{Calliphony}, can be seen as an embodied attempt to extend interaction with real-time generative music models through calligraphy-driven control.
We attach a removable IMU sensor to the brush to capture tri-axial angular velocity during writing, which we map to (1) the per-note inference interval of the model's lead-melody stream (thereby modulating note-onset density) and (2) note-on/off in a chord stream (triggering chord updates, dropping out when still).
The generated MIDI output is then routed to a DAW to drive multiple instrument tracks, enabling a richer and more layered sonic result.

Our contributions are: First, we provide a new perspective on externally controlling real-time symbolic music generation through embodied motion from another artistic modality. 
Second, we introduce a live performance that extends calligraphy beyond a primarily visual art into a multimodal artistic experience.

We will release the code and a video demo upon publication.

\section{Background}

\subsection{Background of Calligraphy}

In Chinese and broader East Asian contexts, calligraphy is commonly understood as a form of visual art that uses written characters as its medium \cite{britannica_chinese_calligraphy}. It involves writing with tools such as a brush and ink—"the art of writing characters with a brush and ink" \cite{oxfordbib_calligraphy}—and organizing the structure and strokes of Chinese characters into aesthetically meaningful "lines and space." This art form has developed alongside writing practices for thousands of years and has established relatively stable learning systems and aesthetic standards, including long-term training in brush technique, character structure, and overall composition, as well as distinct script styles and representative masters. At the cultural level, UNESCO in 2009 inscribed "Chinese calligraphy" on the Representative List of the Intangible Cultural Heritage of Humanity \cite{unesco2009_calligraphy}, and in its decision text emphasized its significance as a symbol of cultural identity and a practice transmitted across generations \cite{unesco_ich_decision_4com_13_08}. Therefore, as a long-standing artistic tradition, calligraphy carries important meanings related to social continuity and cultural identity.

Compared with the cursive or ornamental writing more commonly seen in Western traditions, Chinese calligraphy engages with a comparatively complex logographic writing system: a single character often contains richer structural layers and spatial organization. In addition, the brush's "variable line width" property \cite{calligraphy_interface} means that variations in speed, force, pressing and lifting, and turns are directly reflected in the thickness of ink traces, their dryness or wetness, and an overall sense of rhythm. At the same time, calligraphy is not only about the final static artifact; it also strongly depends on the process of creation. From the perspective of performance and interaction, the act of writing itself has powerful dynamic expressiveness: the writer's posture, the rhythm and continuity of brushwork, and in-the-moment decisions are all visually engaging, and they are well-suited to being translated into real-time multimodal experiences.

\subsection{Calligraphy in NIME and AI-Driven Generative Systems}

Within the NIME community, calligraphy is often framed as an expressive, performative interface for music making. Early systems such as Hé: Calligraphy as a Musical Interface extract computable features from strokes and character shapes (e.g., thickness, length, position) and map them to MIDI parameters, forming a direct "writing-to-sound" interaction loop \cite{calligraphy_interface}. Later work extends this from static feature mapping to the dynamics of the writing process: CalliMusic incorporates timing information (speed, duration, inter-stroke intervals) to shape rhythmic structure, and uses a statistical model to organize note sequences into melodies \cite{calligraphy_brush}. More recent pieces further foreground stage presence and bodily expressivity, treating calligraphic motion itself as performance material—for example, Die Schönheit der Vergänglichkeit interprets nuanced brush behaviors (e.g., speed variation, wet/dry brush) as key musical cues \cite{nime2024_music_20}. Other research shifts the emphasis from system input to collaborative methodology, approaching calligraphy as a "living practice" and positioning the work as documentation and reflection of a shared process\cite{calligraphy_collaboration}.

In the AI era, explorations of calligraphy have also expanded into broader generative and interactive systems. Computer vision and generative modeling have produced substantial work on stylization and controllable generation for Chinese calligraphy—particularly within diffusion-model frameworks—where researchers attempt to generate calligraphy-like visual results under conditions such as character-structure constraints, calligrapher style, and brushstroke features \cite{calliffusion_iccc2023,calliffusionv2_arxiv, moyun_arxiv}. In NIME performance contexts, there are also practices that connect "calligraphic elements – bodily movement – AI systems." For instance, Phantom of Utopia translates calligraphic stroke elements into categories of dancer movement, and uses an AI system to recognize these gestures to trigger audio and visual events, exemplifying a cross-media pathway centered on "AI recognition/triggering" \cite{phantom_utopia}.

Existing NIME work demonstrates that calligraphy's expressivity and performative nature make it well-suited for interactive systems, while AI techniques are often used mainly for sensing and capturing system inputs. However, research that uses calligraphic motion as an external gestural control layer for real-time deep-learning-based symbolic music generation remains relatively rare. In our work, brush motion does not enter the neural model as an additional input feature; instead, it controls when the model is queried, how candidate outputs are constrained, and when additional musical layers are activated.

\subsection{Real-Time Music Generative Models}

Real-time generative music systems have a long history in interactive performance. Earlier systems such as Rowe's interactive music systems, Lewis's \textit{Voyager}, and Pachet's \textit{Continuator} explored how computational agents could listen, respond, continue, or improvise with human performers in real time~\cite{rowe1993interactive,lewis2000voyager,pachet2003continuator}. These works frame generative music not only as offline composition, but also as an interactive process shaped by live human action.

Recent music generative models have expanded this space through stronger statistical modeling of musical structure. Offline systems based on VAEs, diffusion models, Transformers, and language-model architectures have shown increasing ability to generate coherent musical material~\cite{roberts2018musicvae,yang2019deep,mittal2021symbolicdiffusion,min2023polyffusion,wang2024whole,huang2018musictransformer,dhariwal2020jukebox,agostinelli2023musiclm,copet2023musicgen}. However, many of these systems rely on text or audio prompts and are better suited to offline generation than to continuous, low-latency control in live performance.

A key real-time setting is expressive symbolic music performance, where the system produces MIDI-level events incrementally. MIDI event streams can be modeled similarly to token sequences, but they require coordination among multiple musical dimensions, such as pitch, onset, duration, velocity, instrumentation, and inter-part relationships~\cite{Performance,popMAG,MMM,Simon,pop_transformer,Transformer-NADE,notochord}. Online accompaniment systems address one version of this problem by generating harmonically and temporally plausible responses to an unfolding melody~\cite{Bach-Duet,songdriver,realchord}.

For live performance, generative capacity must be balanced with latency and controllability. Recurrent architectures such as RNNs, LSTMs, and GRUs remain useful for event-by-event generation because of their low computational cost, although they are often limited in long-range musical planning. Transformer-based systems can model broader musical context, but may introduce additional latency or interaction-design challenges~\cite{Transformer-NADE,realchord}.

Our work builds on Notochord~\cite{notochord} because its event-level query interface, probabilistic sampling, and multi-instrument support make it suitable for real-time MIDI performance. Rather than treating it as an autonomous composer, we use it as a controllable symbolic generation engine within a larger performance system: calligraphic motion externally controls query timing, candidate constraints, and musical-layer activation.



\section{Methods}

Our technical implementation can be divided into three components: Data Input, Generative Model, and Music Output. These components are handled in Max, Python, and Ableton Live, respectively, and communicate with each other with OSC Protocol\footnote{\url{https://opensoundcontrol.stanford.edu/files/1997-ICMC-OSC.pdf}} for local data transmission.

\begin{figure*}[t]
  \centering
  \includegraphics[width=0.85\textwidth]{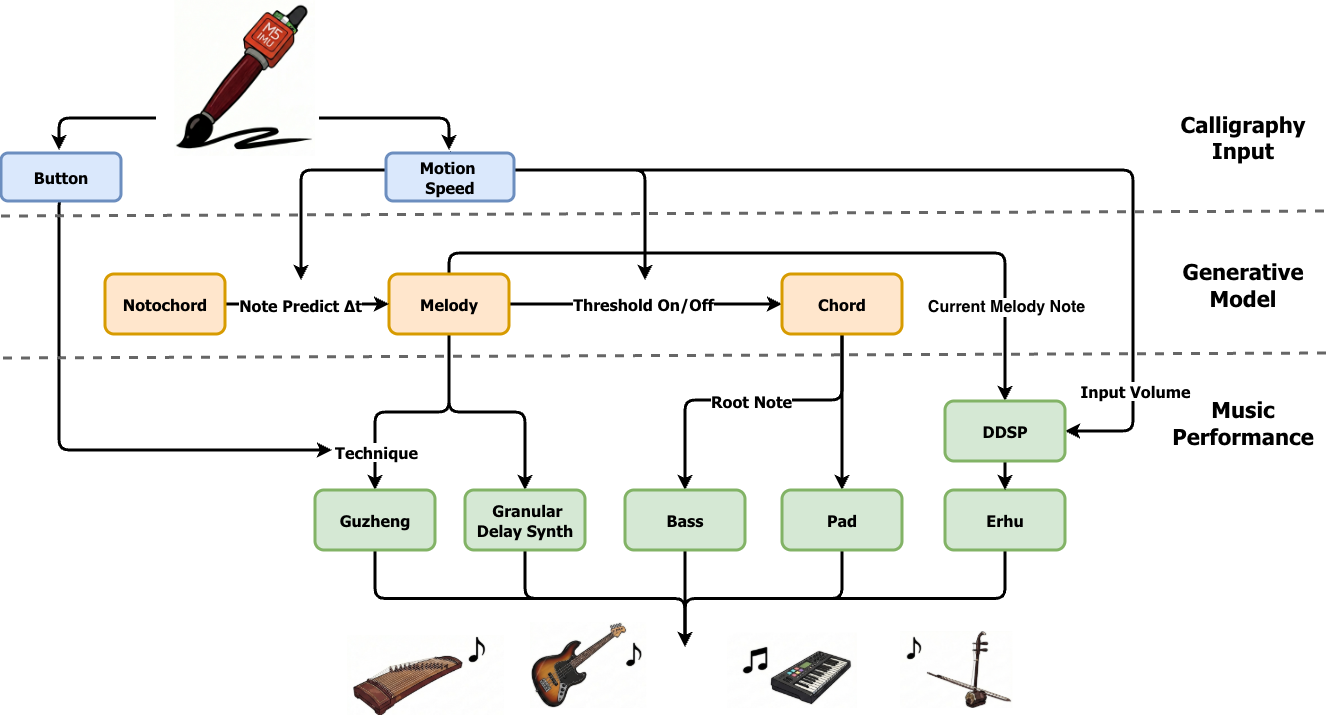}
  \caption{Pipeline of the whole system.}
  \Description{Block diagram of the system pipeline showing data input, generative model, and music output connected via OSC.}
  \label{fig:pipeline}
\end{figure*}

\subsection{Data Input}

We use the built-in gyroscope of an M5Stack\footnote{\url{https://m5stack.com/}} (a modular development board that integrates a microcontroller, a screen, and multiple sensors/expansion interfaces) to capture the brush's rotational motion in space. We chose this brush-mounted gyroscope setup because it preserves the existing writing surface, requires no special paper, and allows the performer to maintain a relatively natural writing posture. Rotational motion also captures salient changes in brush direction and stroke rhythm that are visible in performance. This choice prioritizes portability and stage robustness, while sacrificing information about pressure, contact area, ink flow, and brush-tip deformation.

We wrote an Arduino program so that the M5 sends both the gyroscope data and a trigger signal for each press of button A to a computer via Wi-Fi. In parallel, we 3D-printed a detachable mounting slot that allows the M5 to be fixed onto the brush and quickly removed or replaced when needed. In Max, we process the M5Stack data by accumulating the rotation angles on each axis and normalizing them, obtaining a scalar speed value that represents the overall rotational speed in 3D space. The trigger information from button A is also organized within Max and sent out.

For data routing, we stream this speed value in real time to the Python generation layer, where it controls Notochord's query timing and layer activation. We also send the speed value together with the button-A trigger signal to the third layer, Ableton Live, enabling richer and more fine-grained control of the interactive system.

\begin{figure}[b]
  \centering
  \includegraphics[width=\linewidth]{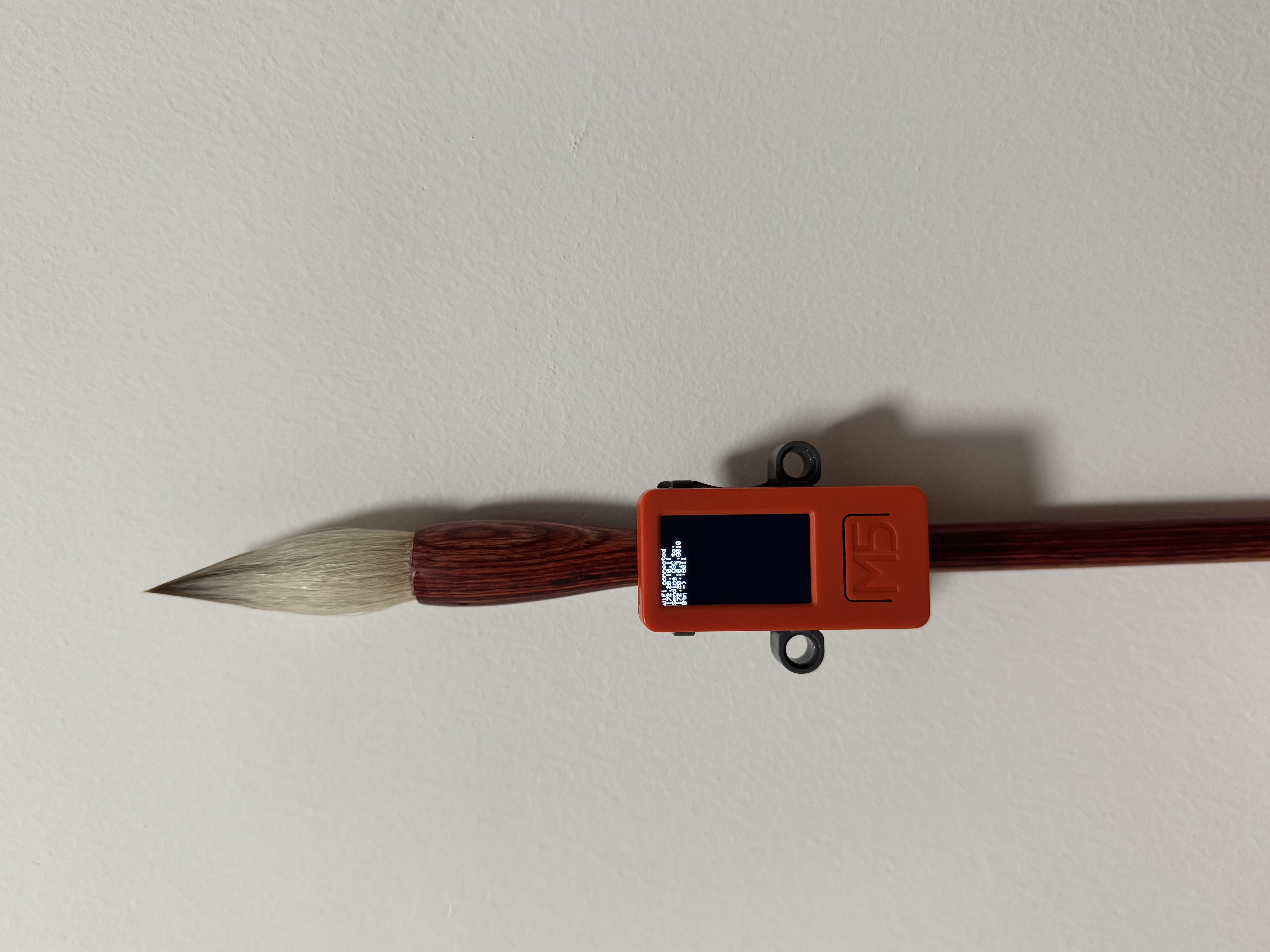}
  \caption{Brush with M5Stack.}
  \label{fig:brush}
\end{figure}

\subsection{Generative Model}
\label{generative model}
The Python layer receives the motion-derived control signals and uses them to control Notochord's inference process. Our system uses the publicly available Notochord model with a modified interaction and inference-control layer. Notochord is an autoregressive neural network model for MIDI performance developed by the Intelligent Instruments Lab \cite{notochord}, designed for real-time interactive music generation in human–machine collaboration. It represents each MIDI event as a combination of four modalities: instrument, pitch, inter-onset interval, and velocity. Instrument and pitch are treated as discrete variables and mapped to a high-dimensional space via embedding layers. Inter-onset interval and velocity are treated as continuous variables and represented using sinusoidal embeddings. The embeddings of the four modalities are summed and fed into a GRU recurrent neural network \cite{gru}, which updates the hidden state event by event to capture temporal dependencies in the performance sequence. We use the publicly released Notochord checkpoint. According to the Notochord release, the model is trained on symbolic MIDI data, including the Lakh MIDI Dataset~\cite{lakh_midi_dataset}.

Notochord provides two core inference interfaces. The feed method inputs an external MIDI event to update the hidden state; the query method samples the next event from the current hidden state under user-specified constraints (e.g., allowed instrument set, pitch range, time truncation, velocity range, sampling temperature, etc.). Building on this, Notochord includes an interactive application called Homunculus, which supports multiple MIDI channels. Each channel can be configured as input mode (input, receiving human performance), follow mode (follow, real-time harmonization), or auto mode (auto, autonomous model generation), with real-time parameter control and a terminal-based text interface. In Notochord's original design, a performer plays in real time on a MIDI keyboard, and the model generates accompaniment parts under given constraints, enabling human–machine collaborative improvisation. However, in our application scenario, we aim to control music generation using external signals unrelated to music (i.e. the rotation of a calligraphy brush).

\subsubsection{Melody}

Based on Notochord's inference interfaces, we designed a new interaction mode. Among the four modalities of each MIDI event, the instrument is fixed to a single timbre; pitch and velocity are still predicted autonomously by the model, but the inter-onset interval is no longer decided by the model. Instead, we integrate an incoming continuous speed signal over time; when the integral value (i.e., accumulated movement distance) reaches a preset threshold, the system triggers the model to predict and output the next note. In this way, the performer's movement speed directly controls the note triggering density: faster motion produces denser notes, while stillness results in silence. We use this mode to control the generation of the main melody.

Notably, in the original Notochord auto mode, notes can end naturally through events with zero velocity. In our system, however, the model only generates note-on (NoteOn) events, while note duration is controlled by an external timer: whenever a new note is triggered, the previous note is terminated immediately, so the system always maintains a monophonic melodic line. The target note duration can be adjusted by the user in real time.

In practical testing, we found that when the model is triggered at a high rate within a short time window, it tends to repeatedly sample the same pitch, resulting in monotonous repeated patterns. To address this, before each query we exclude the most recently played pitch from the candidate set with a certain probability, forcing the model to resample from the remaining pitches. This exclusion probability is adjustable in real time via the interface, ranging from 0 to 1: at 0, no restriction is applied; at 1, the previous pitch is always excluded.

For pitch constraints, the original Notochord does not enforce any scale, and the model may predict any semitone. To make the generated melody conform to a specific tonality, we apply scale-based filtering to the pitch candidate set. The system provides multiple built-in scale modes—such as major, minor, pentatonic, and blues—and the model is only allowed to sample from the pitch set corresponding to the currently selected scale. Meanwhile, the user can transpose the melody in real time using octave and semitone offset parameters to adjust the register.

In addition, to increase the naturalness and expressivity of the generated music, we introduce several real-time adjustable randomization parameters. Gaussian noise can be added to the trigger threshold, velocity, and note duration, making the resulting sound more human-like.

\begin{figure}[t]
  \centering
  \includegraphics[width=0.9\linewidth]{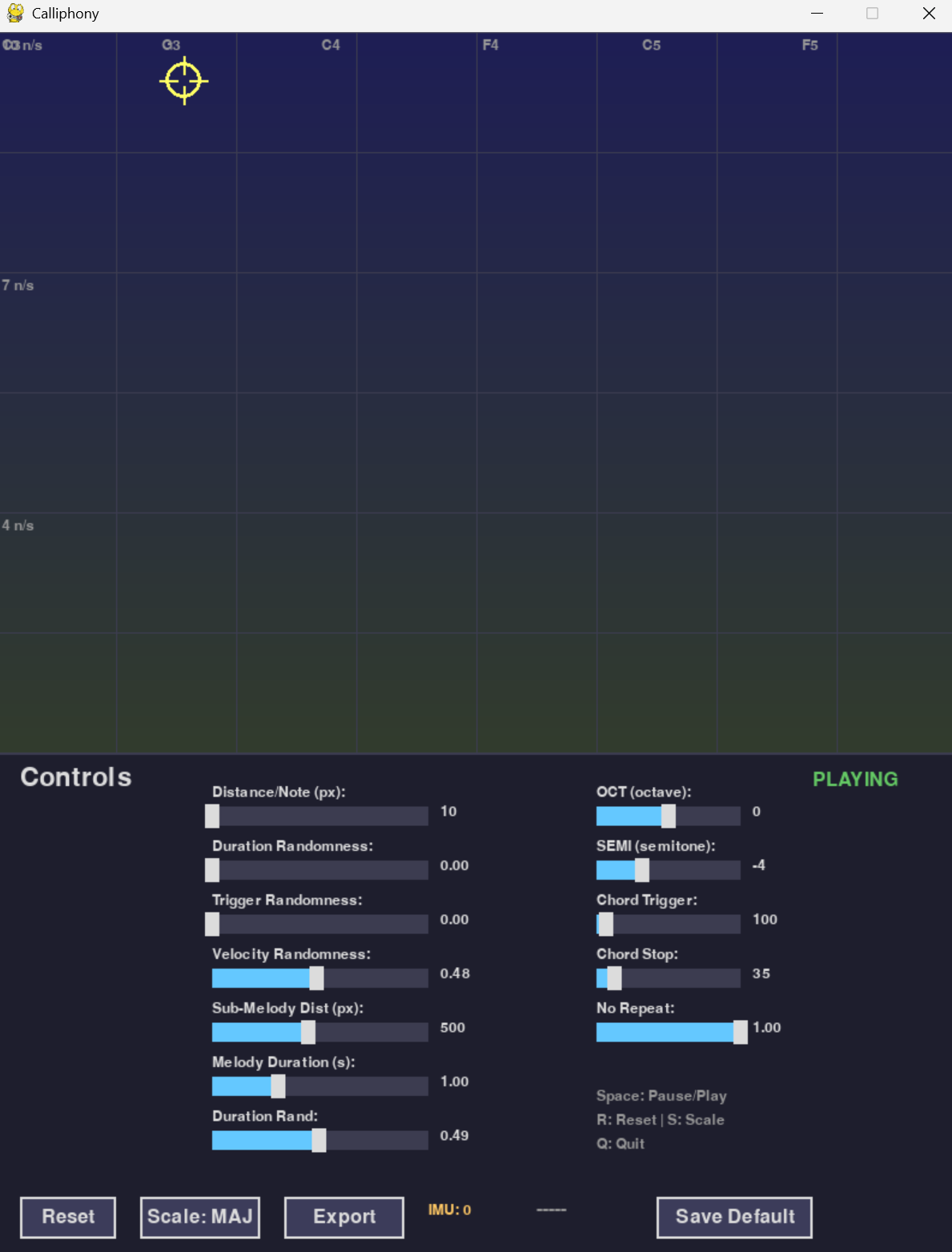}
  \caption{Control UI for real-time parameter tuning and note-trigger visualization.}
  \label{fig:GUI}
\end{figure}

\subsubsection{Chord, Bass, and Sub-melody}

In addition to the main melody, we designed a speed-driven coupling mechanism for chords, bass, and a sub-melody, each routed to an independent MIDI channel. Unlike the melody, which is triggered by distance integration, these three parts use threshold-based triggering: when the movement speed exceeds a user-defined onset threshold, all three parts start simultaneously; when the speed falls below a stop threshold, all three parts are silenced simultaneously. Both the onset and stop thresholds can be adjusted in real time. The difference between the two thresholds creates a hysteresis band, preventing repeated triggering and releasing near the threshold.

Chord generation uses Notochord's query interface under externally specified constraints. We treat the current main-melody pitch as a provisional root and sample two additional harmony notes through two constrained query calls, with candidate pitches limited to the selected scale and to a register around the root. During each sampling step, any pitch that has already been selected is removed from the candidate set, ensuring that the three chord tones are not duplicated.

For the bass part, we take the chord root and transpose it down by one octave, then constrain it to a fixed low register (an octave range centered on C2), and output it to a dedicated MIDI channel. For the sub-melody, we take the current main-melody pitch and constrain it to a mid register (an octave range centered on C4). Its velocity is obtained by applying a moving-average smoothing to the current movement speed and then mapping it linearly: faster motion produces stronger sub-melody velocity.

With this design, the performer's movement speed controls not only the density of melodic note onsets, but also the entrance and exit of the harmonic parts: with slow movement, only a monophonic melody is present; as the performer accelerates, chords, bass, and the sub-melody are added automatically; as the performer slows down, these accompaniment parts disappear accordingly. The interleaving between the melodic and harmonic tracks makes the generated music more layered.

The system provides a graphical control interface developed in Pygame\footnote{\url{https://www.pygame.org}}. Users can adjust the above parameters in real time via sliders (e.g., trigger threshold, scale mode, transposition, exclusion probability, chord trigger threshold), and the interface also displays real-time visual feedback of note triggering. During development and debugging, the system supports using mouse movement to simulate motion input for testing; in actual performances, the motion signals are received from an external source.

\subsection{Music Output}

The MIDI output from the generative model is routed in real time to Ableton Live via a LoopMIDI virtual port, with different melodic and harmonic tracks played using different timbres. To align with Chinese calligraphy as a culturally distinctive symbol, we also favor traditional Chinese instruments in our instrument choices across tracks.

The main-melody track layers a guzheng (a traditional Chinese plucked string instrument) sampled instrument with a synthesizer processed by a Granular Delay effect. The guzheng sound uses a virtual instrument developed by Cat Audio\footnote{\url{https://cataudio.cn/}}; it supports switching among four performance techniques—thumb, tremolo (yaozhi), light vibrato, and harmonics—via the A button on the M5Stack. The chord track uses a pad timbre. The sub-melody track uses a DDSP timbre-transfer model based on prior work on gesture-driven erhu synthesis \cite{DDSPerhu}. This track takes an electronic synthesizer timbre as input and outputs an erhu timbre after timbre transfer through the DDSP model. We map the movement-speed signal (after smoothing) to the track volume. The bass track uses a Reese Bass synthesizer timbre, and its volume is likewise controlled by the speed signal.

In performance, the system produces a broad correspondence between calligraphic energy and musical density. Slow brush movement tends to produce sparse monophonic melodic fragments, while faster movement increases note-onset density. When the speed exceeds the accompaniment threshold, the chord, bass, and sub-melody layers enter together, creating a thicker texture. When the performer slows down or pauses, these layers drop out, returning the music to a sparser state or silence. The guzheng layer gives the main melody a plucked and articulated quality; the pad layer provides harmonic background; the DDSP-based erhu timbre introduces a bowed-instrument color; and the bass layer reinforces high-energy moments.

\section{Performance Demonstration and Feedback}

After completing the system design, the author gave an approximately five-minute live performance at an art exchange event. The performance took place in a studio: the performer held a calligraphy brush equipped with an M5Stack sensor and wrote the well-known Chinese classical poem \textit{Shui Diao Ge Tou} on a water-writing cloth.

Participants in the event were researchers and practitioners in the arts. Most had grown up in an East Asian cultural context and had some familiarity with calligraphy. After the performance, we collected informal audience feedback. Some audience members showed strong interest in the interaction and generation of the sound; they focused on the mapping relationships between different musical tracks and the writing gestures, and wanted to further understand the technical principles of real-time interaction between the music generation system and the performer's movements. Others were more concerned with the data acquisition during writing—for example, whether different stroke shapes and brush pressure influenced the generated music. Since the performance was not formally recorded and no systematic questionnaire was conducted, these comments do not constitute user-study data in a strict sense, but they still provided valuable directions for subsequent design iterations.

\begin{figure}[t]
  \centering
  \includegraphics[width=\linewidth]{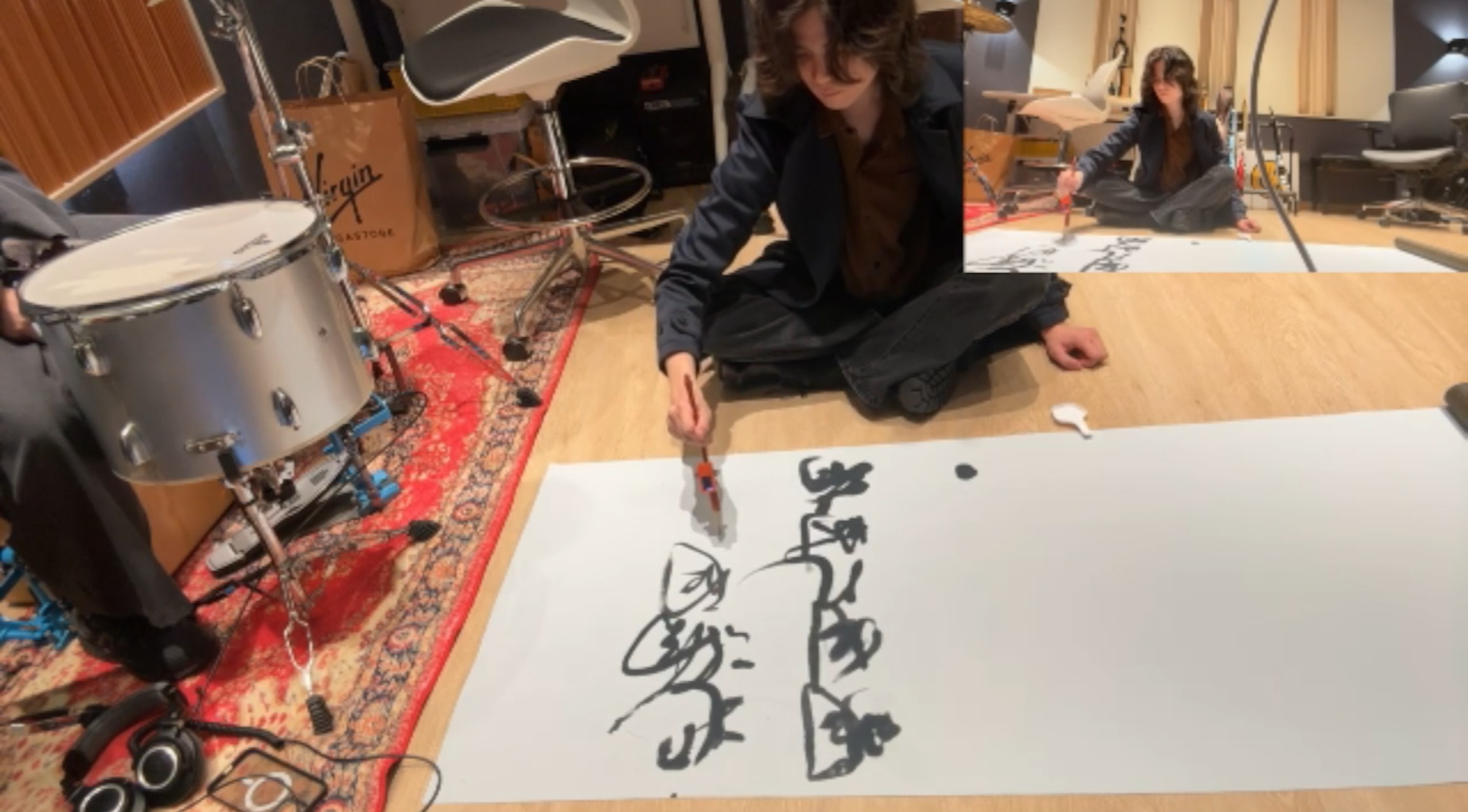}
  \caption{Live performance of Calliphony.}
  \label{fig:performance}
\end{figure}

\section{Discussion}


Notochord's GRU-based recurrent architecture \cite{gru} has inherent limitations in modeling long-range dependencies. While it can produce plausible note sequences within a local context, it struggles to form clear phrase structures or sectional hierarchies over longer time scales, and the resulting music tends to lack macro-level organization.

On the input side, the current system relies only on the M5Stack's built-in gyroscope to capture motion during writing, which limits the dimensionality of the sensed information. For instance, subtle pressure variations in brush handling and the contact area between the brush tip and the writing surface—both crucial to calligraphic expressivity—cannot be effectively captured with the current hardware.

Therefore, to improve real-time generative performance in similar scenarios, a dedicated MIDI dataset for Chinese traditional performing arts is especially important. Such a dataset would enable generative models to learn melodic and harmonic patterns that better match the aesthetic requirements of the target context. At the model level, future work could explore architectures with stronger long-range modeling capabilities and investigate the trade-off between computational cost and real-time generation constraints. At the input level, richer sensing modalities could be introduced to capture multidimensional information in the writing process—for example, using pressure sensors on the writing surface to measure brush force, or using surface electromyography (sEMG) to sense muscle activity in the arm and wrist—thereby providing finer-grained control signals for music generation.

\section{Conclusion}

Starting from Chinese calligraphy as an East Asian intangible cultural heritage, this work shows how AI can extend traditional arts into new digital performance settings. We frame calligraphy as an embodied, time-based practice, and present a real-time cross-modal system that uses writing gestures to externally control symbolic music generation, allowing brush dynamics to be rendered as changes in musical density, texture, and timbre.

We implement a layered Max/Python/Ableton Live architecture with M5Stack motion sensing and OSC control. Brush speed is translated into two core control mechanisms: distance-integral triggering for lead-melody events, and threshold-based activation/deactivation for chords, bass, and a sub-melody. Additional constraints—scale filtering, pitch-repeat suppression, and adjustable randomness—improve musical coherence while preserving responsiveness for live performance.

Beyond a specific system, this work contributes a practical paradigm for AI-assisted co-creation with traditional arts: using embodied motion as an external control layer for real-time generative systems, and using cross-modal translation to connect expressive processes across media.

\section{Acknowledgments}

We thank Daniel for support with 3D printing, Ruohan for valuable discussions on combining AI and calligraphy, and Gus for guidance on writing.
\section{Ethical Standards}

This work involved no human-subject study and collected no personally identifiable information. The system was used in a public artistic demonstration with informal, voluntary feedback.

\bibliographystyle{plain} 
\bibliography{references} 
\end{document}